\documentclass[fleqn,usenatbib]{mnras}

\usepackage{newtxtext,newtxmath}

\usepackage[T1]{fontenc}
\usepackage{ae,aecompl}

\usepackage{graphicx}	
\usepackage{amsmath}	
\usepackage{bm}
\usepackage{dblfloatfix}
\usepackage{ulem}

\usepackage{scalerel,tikz}
\usetikzlibrary{svg.path}
\definecolor{orcidlogocol}{HTML}{A6CE39}
\tikzset{orcidlogo/.pic={
 \fill[orcidlogocol] svg{M256,128c0,70.7-57.3,128-128,128C57.3,256,0,198.7,0,128C0,57.3,57.3,0,128,0C198.7,0,256,57.3,256,128z};
 \fill[white] svg{M86.3,186.2H70.9V79.1h15.4v48.4V186.2z}
 svg{M108.9,79.1h41.6c39.6,0,57,28.3,57,53.6c0,27.5-21.5,53.6-56.8,53.6h-41.8V79.1z M124.3,172.4h24.5c34.9,0,42.9-26.5,42.9-39.7c0-21.5-13.7-39.7-43.7-39.7h-23.7V172.4z}
 svg{M88.7,56.8c0,5.5-4.5,10.1-10.1,10.1c-5.6,0-10.1-4.6-10.1-10.1c0-5.6,4.5-10.1,10.1-10.1C84.2,46.7,88.7,51.3,88.7,56.8z};
}}
\newcommand\orcidicon[1]{\href{https://orcid.org/#1}{\mbox{\scalerel*{
\begin{tikzpicture}[yscale=-1,transform shape]
\pic{orcidlogo};
\end{tikzpicture}
}{|}}}}

\title[HMF in VVV]{The abundance of dark matter haloes down to Earth mass}
\author[H. Zheng et al.]{%
Haonan Zheng\orcidicon{0000-0002-1665-5138}$^{1,2,3}$\thanks{Email: hnzheng@nao.cas.cn}, 
Sownak Bose\orcidicon{0000-0002-0974-5266}$^{3}$, 
Carlos S. Frenk\orcidicon{0000-0002-2338-716X}$^{3}$, 
Liang Gao\orcidicon{0000-0002-9276-917X}$^{1,2,4,5}$\thanks{Email: lgao@bao.ac.cn}, 
Adrian Jenkins\orcidicon{0000-0003-4389-2232}$^{3}$, 
\newauthor 
Shihong Liao\orcidicon{0000-0001-7075-6098}$^{1}$, 
Yizhou Liu\orcidicon{0009-0005-8855-0748}$^{1,2}$, 
Jie Wang$^{1,2,4}$
\vspace*{0.1cm}\\%
$^{1}$Key Laboratory for Computational Astrophysics, National Astronomical Observatories, Chinese Academy of Sciences, Beijing 100101, China\\%
$^{2}$School of Astronomy and Space Science, University of Chinese Academy of Sciences, Beijing 100049, China\\%
$^{3}$Institute for Computational Cosmology, Department of Physics, University of Durham, South Road, Durham, DH1 3LE, UK\\%
$^{4}$Institute for Frontiers in Astronomy and Astrophysics, Beijing Normal University, Beijing 102206, China\\%
$^{5}$School of Physics and Microelectronics, Zhengzhou University, Zhengzhou 450001, China\\%
}

\date{Accepted XXX. Received YYY; in original form ZZZ}

\pubyear{2024}

\begin{document}
\label{firstpage}
\pagerange{\pageref{firstpage}--\pageref{lastpage}}
\maketitle

\begin{abstract}
  We use the Voids-within-Voids-within-Voids (VVV) simulations, a suite
of successive nested N-body simulations with extremely high resolution
(denoted, from low to high resolution, by L0 to L7), to test the
Press-Schechter (PS), Sheth-Tormen (ST), and extended Press-Schechter
(EPS) formulae for the halo abundance over the entire mass range, from
mini-haloes of $10^{-6}\ \mathrm{M_\odot}$, to cluster haloes of
$10^{15}\ \mathrm{M_\odot}$, at different redshifts, from $z=30$ to
the present.  We find that at $z=0$ and $z=2$, ST best reproduces the
results of L0, which has the mean cosmic density (overdensity
$\delta=0$), at $10^{11-15} ~\mathrm{M_\odot}$. The higher resolution
levels (L1-L7) are biased underdense regions ($\delta<-0.6$). The EPS
formalism takes this into account since it gives the mass function of
a region conditioned, in this case, on having a given
underdensity. EPS provides good matches to these higher levels, with
deviations $\lesssim 20\%$, at $10^{-6-12.5} ~\mathrm{M_\odot}$.  At
$z \sim 7-15$, the ST predictions for L0 and the EPS for L1-L7 show
somewhat larger deviations from the simulation results. However, at
even higher redshifts, $z \sim 30$, EPS fits the simulations well
again.  We confirm our results by picking more subvolumes from the L0
simulation, finding that our conclusions depend only weakly on the
size and overdensity of the region.  The good agreement of EPS with
the higher-level simulations implies that PS (or ST)
gives an accurate description of the total halo mass function in
representative regions of the universe.

\end{abstract}




\section{Introduction}
\label{sec:intro}

In the Lambda Cold Dark Matter ($\Lambda$CDM) model structures grow
hierarchically from primordial quantum fluctuations to the galactic
haloes and large-scale structures in the universe observed today
\citep{Davis1985}. The abundance of these structures as a function of
their mass provides a fundamental basis for galaxy formation models
\citep{White&Frenk1991} and a framework for constraining cosmological
parameters \citep{Frenk1990, White1993, Henry2009, 
  Allen2011ARA&A}. It is important, therefore, to develop theoretical
models that describe how initial density perturbations collapse into
non-linear structures, and to use these as tools for interpreting the
predictions of N-body simulations.

Using the statistics of Gaussian random fields and the spherical
collapse model, \citet{PS1974} proposed the well-known analytical
model for halo mass functions, the so-called Press-Schechter formalism
(PS model hereafter), which, at least qualitatively, predicts halo
abundances that are comparable to numerical simulations. However, the
PS model overpredicts the halo mass function in the low-mass regime
(by even up to $60\%$ at $z=0$) and exhibits too sharp a decrease at
the high-mass end \citep{White1993, Gross1998, Governato1999,
  Jenkins2001, Lukic2007, Pillepich2010}. To remedy this,
\citet{ST1999, ST2002} replaced the spherical collapse ansatz in the
PS model with the ellipsoidal collapse model and provided another
formula (ST model hereafter) whose predictions provide a somewhat
closer match to those of numerical simulations. For example, according
to \citet{Reed2003}, the difference between the ST halo mass function
and their simulated one is $\lesssim 10\%$ in well-sampled mass bins.

\cite{Bond1991}, \cite{Bower1991} and \cite{Lacey1993} extended the PS model (hereafter
the EPS model), to include predictions for the halo abundance and
assembly history in different environments \citep{Gao2005,
  Faltenbacher2010}. Thereafter, several studies have attempted to
provide accurate and universal fitting formulae for halo mass functions
from numerical simulations \citep[e.g.][]{Jenkins2001, Warren2006,
  Reed2007, Lukic2007, Tinker2008, Crocce2010, Manera2010, Watson2013,
  Despali2016}, improving upon earlier fits by extending to higher redshifts
and wider mass ranges, and also considering different definitions of
halo mass.

It is challenging to perform cosmological simulations with high mass
resolution over a wide mass range down to redshift $z=0$ because of
computational cost. Previous studies, therefore, have focused
primarily on either high-mass haloes
($10^{10-15}~\mathrm{M_\odot}$) down to low redshift
\citep[e.g.][]{Jenkins2001, Tinker2008} or low-mass haloes
($10^{5-10}~\mathrm{M_\odot}$) at high redshifts \citep[e.g.
$z=10$, $30$, see][]{Lukic2007, Reed2007}. As a result, there are
comparatively few studies comparing the theoretical halo mass
functions (e.g. the PS, ST, and EPS models) to simulations at the mass
range of mini-haloes ($\lesssim 10^5~\mathrm{M_\odot}$) down to low
redshift \citep{Angulo2017}. Furthermore, most comparisons of EPS
predictions with simulations focus on overdense or mean density
regions.
In this work, we use the VVV simulations \citep{Wang2020}, a series of
successive nested zoom cosmological simulations of void regions with
extremely high mass resolution, to test the accuracy of theoretical
halo mass functions over the full mass range of CDM haloes as a
function of time. By construction, this work focuses on the abundance
of haloes in preferentially underdense regions.

The paper is organized as follows. Section~\ref{sec:theory} briefly
describes the PS, ST, and EPS theoretical halo mass functions and
Section~\ref{sec:simulation} the details of our simulations. Our
results, discussions and conclusions are presented in Sections 
\ref{sec:results}, \ref{sec:discussion}, 
and \ref{sec:conclusion}, respectively. An examination of the
conversion between linear and non-linear overdensities are illustrated
in Appendix~\ref{ap:eq6}. Further tests with the EAGLE simulations
\citep{Schaye2015} are presented in Appendix \ref{ap:eagle34}.

\section{Theoretical halo mass functions}
\label{sec:theory}

Press-Schechter theory \citep{PS1974} assumes that the initial density
field follows a Gaussian random distribution, and that gravitational
collapse occurs when the smoothed density field, $\delta$, exceeds the
critical overdensity for collapse, $\delta_\mathrm{c}$, during its
random walk in the space of $\sigma(M)-\delta$, where $\sigma(M)$
represents the variance of the density field smoothed with a filter of
mass, $M$, by the conventional real-space top-hat filter \citep[e.g.][]{Lacey1993, Mo2002}. 
These crossing events correspond to structure formation on a
certain scale, yielding a halo mass function given by:
\begin{equation}
    n_{\mathrm{PS}}(M, z)\mathrm{d} M = \sqrt{\frac 2 \pi} \frac{\bar{\rho}_0}{M}\frac{\mathrm{d}\nu}{\mathrm{d}M}\exp \left(-\frac{\nu^2}{2}\right)\mathrm{d}M, 
\end{equation}
where $\bar{\rho}_0 = \Omega_\mathrm{m}\rho_{\mathrm{crit,}z\mathrm{=0}}$ is the mean matter density, $\nu(M, z) = \delta_\mathrm{c} / [D(z) \sigma(M)]$, $\delta_\mathrm{c} = 1.68647$, 
and $D(z)$ is the growth factor given by $D(z)=g(z)/\left[g(0)(1+z)\right]$, with (assuming a flat universe, i.e. $\Omega_\mathrm{m} + \Omega_\Lambda = 1$):
\begin{equation}
    g(z) = a E(z)\int^{a}_{0}\left(\Omega_\mathrm{m} x^{-1} + \Omega_\Lambda x^2\right)^{-\frac 3 2} \mathrm{d}x,
\end{equation}
\vspace{-0.1cm}
\begin{equation}
    E(z) = \frac{H(z)}{H_0} = \left(\Omega_\mathrm{m} a^{-3} + \Omega_\Lambda\right)^{\frac 1 2},
\end{equation}
and $a$ is the scale factor defined as $(1+z)^{-1}$. 

While PS theory considers a spherical model for the collapse of
perturbations, the Sheth \& Tormen \citep{ST1999, ST2002} model takes
ellipsoidal collapse\footnote{The spherical
  collapse model considers the evolution of a spherically symmetric
  density perturbation; the ellipsoidal model makes the more
  realistic assumption that the overdense region is an ellipsoid,
  which leads to different collapse times in different directions.}
  into consideration, leading to a modified formula: 
\begin{equation}
    n_{\mathrm{ST}}(M, z)\mathrm{d} M = A_\mathrm{ST}\left(1+\frac{1}{\nu'^{2q_\mathrm{ST}}}\right)\sqrt{\frac 2 \pi} \frac{\bar{\rho}_0}{M}\frac{\mathrm{d}\nu'}{\mathrm{d}M}\exp \left(-\frac{\nu'^2}{2}\right)\mathrm{d}M,
\end{equation}
where $\nu'=\sqrt{a_\mathrm{ST}} \nu$, with free parameters $a_\mathrm{ST} = 0.707$, $A_\mathrm{ST}=0.322$ and $q_\mathrm{ST}=0.3$. 

Extended PS theory \citep{Bond1991, Lacey1993, Mo1996} is a useful
tool to quantify halo assembly history and assembly bias. In
particular, it predicts the probability that matter in a spherical
region of mass, $M_0$, at redshift, $z_0$, and linear overdensity,
$\delta_0$, is contained within dark matter haloes of mass in
the interval $(M_1, M_1 + \mathrm{d}M_1)$ at redshift $z_1$:
\begin{equation}
    \begin{split}
        f(M_1, \delta_1 | M_0, \delta_0) \mathrm{d}M_1 & = \sqrt{\frac{1}{2\pi}} \frac{\delta_1 - \delta_0}{(\sigma^2_1 - \sigma^2_0)^{\frac 3 2}}\times \\
        & \quad \exp \left[-\frac{(\delta_1 - \delta_0)^2}{2(\sigma^2_1 - \sigma^2_0)} \right] \frac{\mathrm{d} \sigma^2_1}{\mathrm{d} M_1} \mathrm{d}M_1,
    \end{split}
  \end{equation}
  where, for a virialized structure, $\delta_1=\delta_\mathrm{c}/D(z)$,
and $\delta_0$ is given by \footnote{The
  accuracy of this formula in underdense regions is examined in
  Appendix~\ref{ap:eq6}.}: 
\begin{equation}
    \label{eq:6}
    \begin{split}
        \delta_0(\delta_\mathrm{nl}, z) &= \frac{\delta_\mathrm{c} / D(z)}{1.68647} \times C(\delta_\mathrm{nl}) [1.68647 - 1.35(1+\delta_\mathrm{nl})^{-\frac 2 3} - \\ 
        & \quad 1.12431 (1+\delta_\mathrm{nl})^{-\frac 1 2} + 0.78785 (1+\delta_\mathrm{nl})^{-0.58661} ],
    \end{split}
\end{equation}
as an approximate solution of Eqs.~(16-17) of \citet{Mo1996}. Equation~\ref{eq:6} is just an approximation
to the exact solution that relates the nonlinear density  to the linear density during the collapse of a homogeneous sphere.
Here, 
$\delta_\mathrm{nl}$ denotes the non-linear overdensity in Eulerian
space, which can be calculated directly from an N-body simulation. We modified
the original formula from \citet{Mo1996, ST2002} by including an additional
factor, $C(\delta_\mathrm{nl})$, to improve the accuracy of the fit when the
non-linear overdensity is close to -1. We carried out our own fit to the spherical
collapse model and obtained a correction factor of  $C(\delta_\mathrm{nl})=1-0.0053977x+0.00184835x^2+0.00011834x^3$ with $x=\mathrm{min}(0, \ln{(1+\delta_\mathrm{nl})})$.  This makes a difference of about 5\% for  $\delta_\mathrm{nl} = -0.99$. 

Once these are calculated, we may obtain the corresponding halo mass
function as: 
\begin{equation}
    n_{\mathrm{EPS}}(M, z)\mathrm{d} M = \frac{(1 + \delta_\mathrm{nl})\bar{\rho}_0}{M}f(M, \delta_1 | M_0, \delta_0)\mathrm{d}M,
\end{equation}
where a factor of $(1 + \delta_\mathrm{nl})$ is introduced to accommodate the difference of the sphere sizes in Eulerian and Lagrangian space. 
\begin{figure*}
    \centering
    \vspace{-0.25cm}
	\includegraphics[width=2.0\columnwidth]{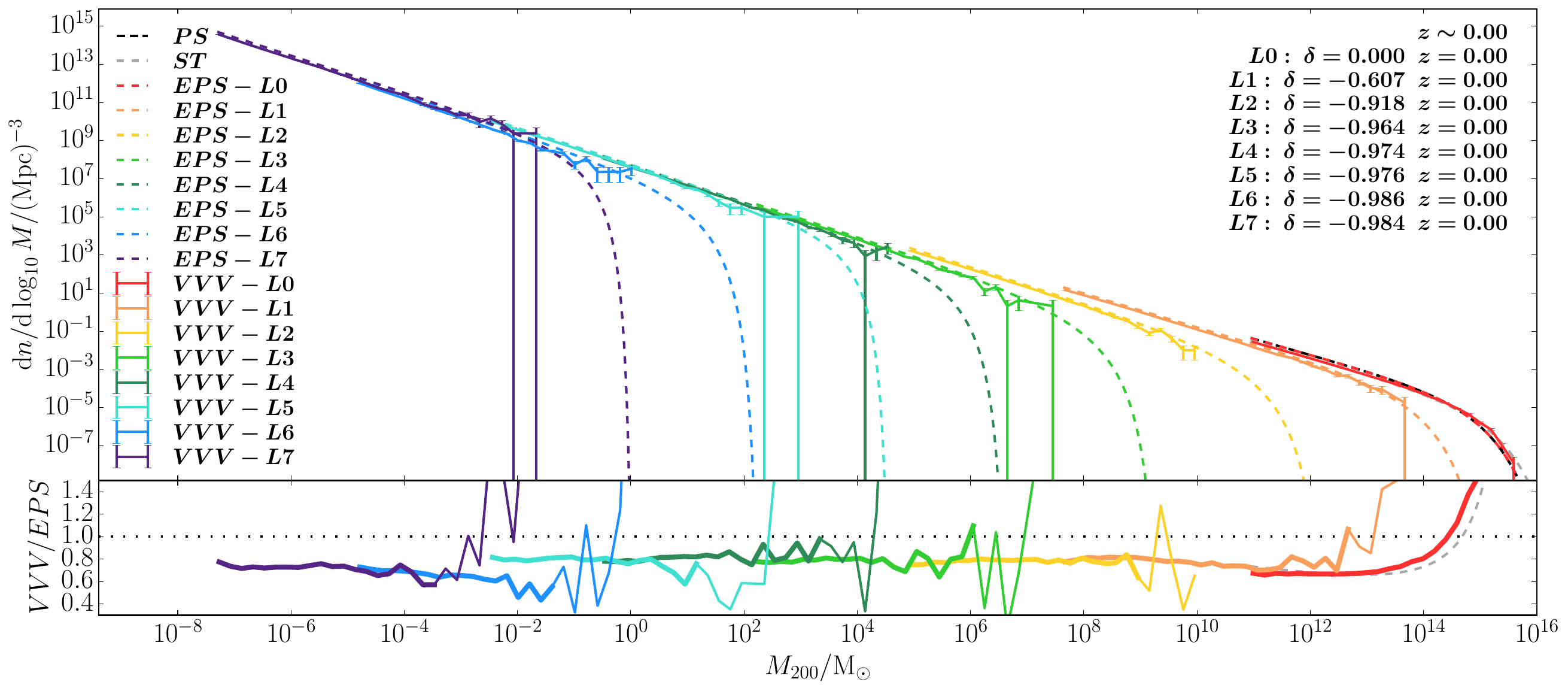}
	\vspace{-0.30cm}
	\caption{Predicted and simulated halo mass functions in
          different resolution levels at $z=0$. In the top panel, the
          colour solid lines show the halo mass functions in the VVV (error
          bars represent Poisson errors); black, gray and colour 
          dashed lines  show predictions of
          the PS, ST, and EPS models for each resolution level
          respectively. In the bottom panel,
          the solid lines show the ratio VVV/EPS.  Thick lines
          indicate mass bins containing at least 20 haloes. The
          black dotted line represents a ratio of unity, while the gray dashed line
          represents the ratio ST/PS. At $z=0$, ST gives the best
          prediction for L0; EPS  overpredicts the halo abundance
          in higher resolution levels by $\sim 20\%$. }
	\vspace{-0.20cm}
	\label{fig:fig1}
	
	
\end{figure*}
It is worth noting that, in the limit where the density of a large
region is equal to the cosmic mean matter density (i.e.,
$\delta_\mathrm{nl}=0$), the EPS formula reverts to the standard PS
formula.  

\begin{figure*}
    \centering
	\includegraphics[width=2.0\columnwidth]{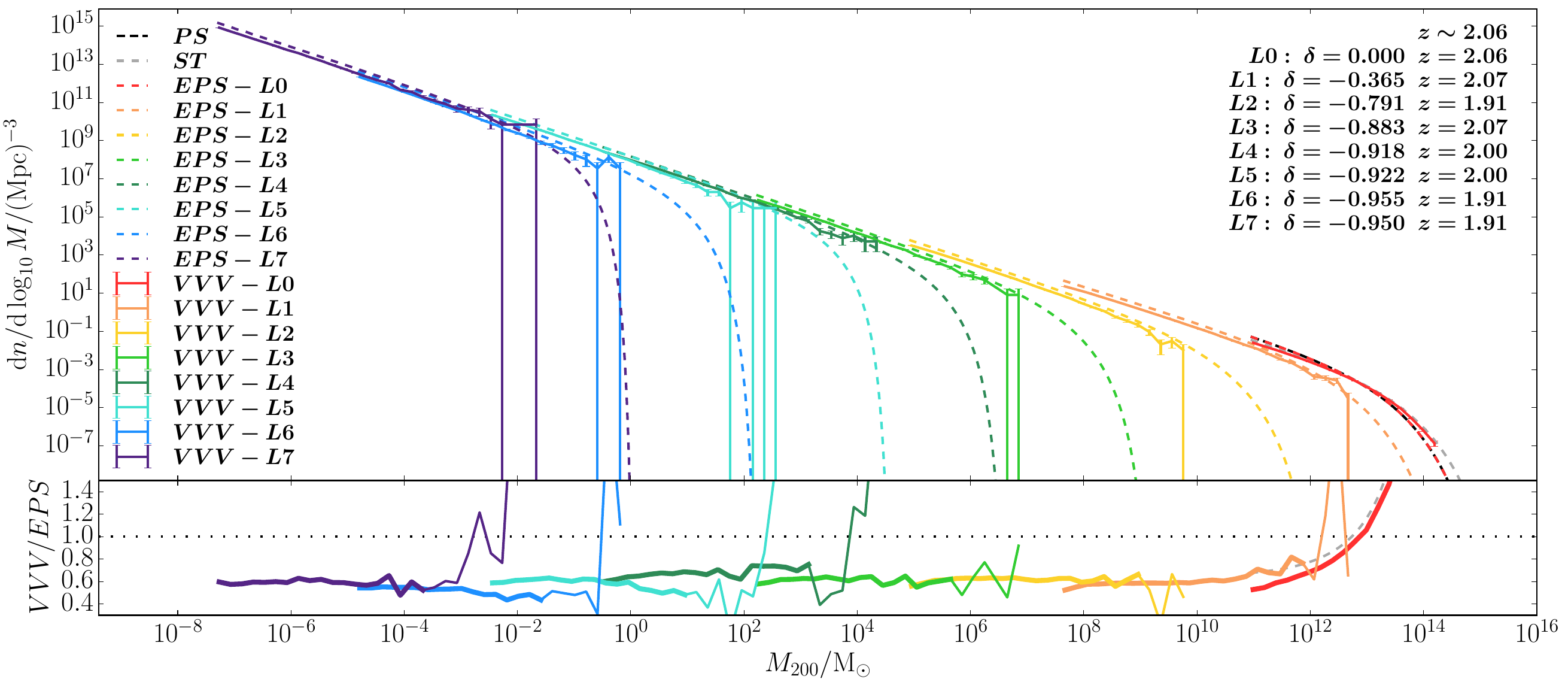}
	\vspace{-0.00cm}
	\includegraphics[width=2.0\columnwidth]{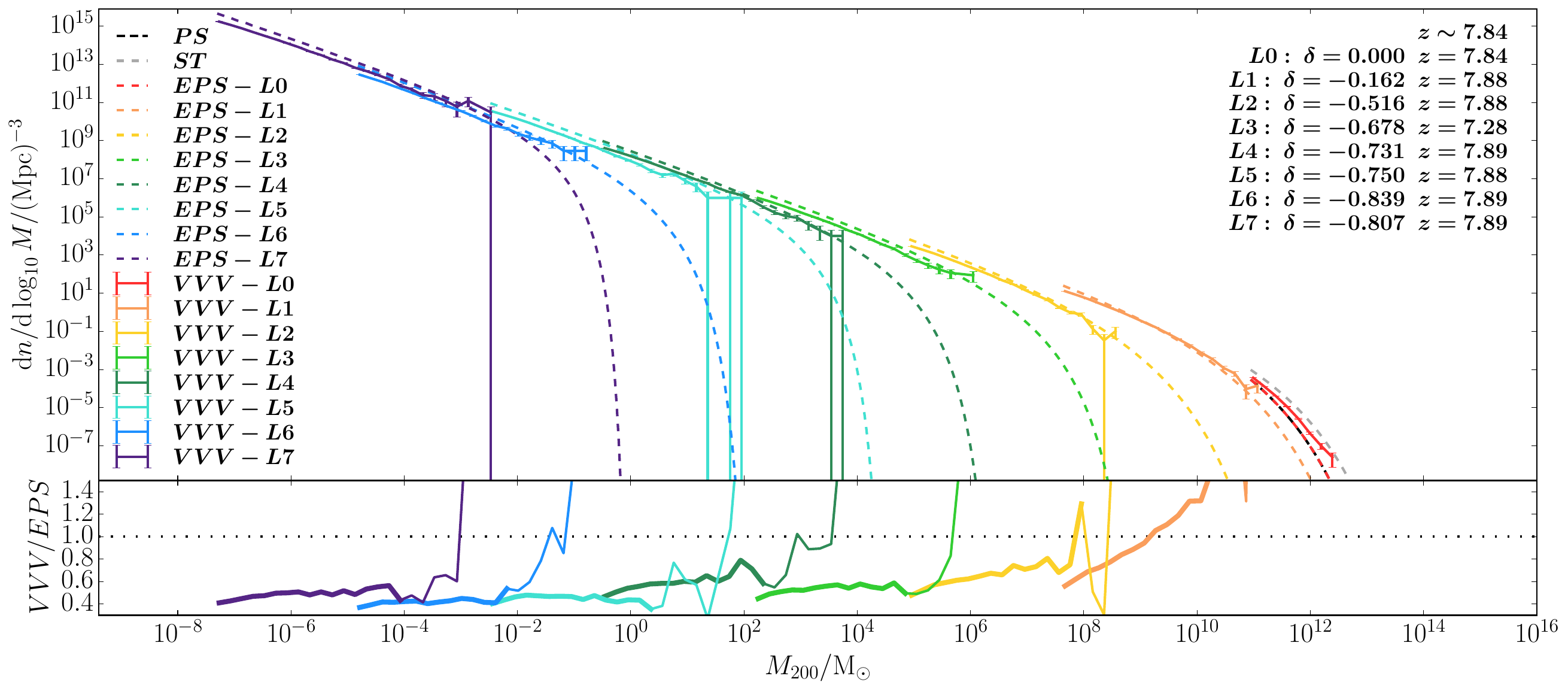}
	\vspace{-0.00cm}
	\includegraphics[width=2.0\columnwidth]{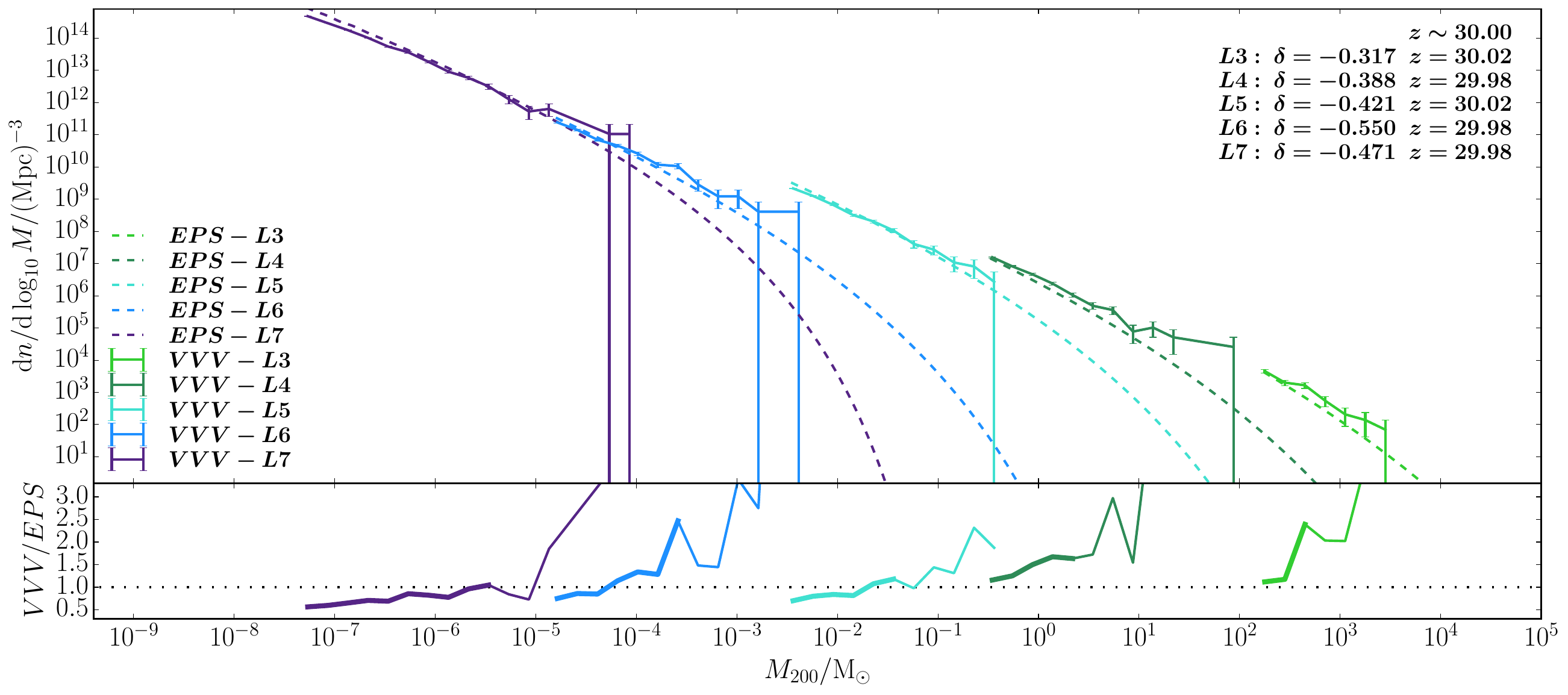}
	\vspace{-0.30cm}
	\caption{As Fig.~\ref{fig:fig1}, but for $z \sim 2$ (top
          panel), $z\sim 7.8$ (middle panel), and $z\sim 30$ (bottom
          panel). For $z \sim 30$, we only show results from L3-L7 as
          very few haloes have formed at this time in the lower
          resolution levels. The PS and ST predictions are calculated
          at the corresponding redshifts, while the simulation results
          and the EPS predictions are shown at the closest available
          redshift output in the simulation. At higher redshifts, the
          EPS prediction deviates more from the simulations, by
          $\sim 20-50\%$, especially at $z \sim 7.8$. }
	
	\label{fig:fig2}
\end{figure*}

\section{Details of the simulations}
\label{sec:simulation} 

We use the nested zoom N-body simulations of the VVV project
\citep{Wang2020}, which consists of 8 levels of resimulation, covering
a wide halo mass range spanning around 20 orders of magnitude
($\sim 10^{-6}-10^{15}~\mathrm{M_\odot}$). These simulations were
performed with the \textsc{Gadget-4} code \citep{Springel2020},
adopting cosmological parameters derived from Planck
\citep{Planck2014p16}: $\Omega_{\rm{m}} = 0.307$,
$\Omega_{\Lambda} = 0.693$, $h = 0.6777$, $n_\mathrm{s} = 0.961$ and
$\sigma_8 = 0.829$. On large scales ($k \le 7~\mathrm{Mpc}^{-1}$), the
initial linear power spectrum is computed with the \textsc{Camb} code
\citep{Lewis2000}, while the BBKS fitting formula \citep{BBKS1986}
with $\Gamma=0.1673$ and $\sigma_8=0.8811$ is adopted to extrapolate
the power spectrum to small scales ($k \ge 70~\mathrm{Mpc}^{-1}$),
with a smooth transition between $7-70~\mathrm{Mpc}^{-1}$.

\begin{table}
 \caption{Properties of the different resolution levels in the VVV simulation
   used in this work. Column 1: name of the level; column 2:
   size ($L_{\mathrm{box}}$ or $d_{\mathrm{sphere}}$) of the region 
   selected at each resolution  level at $z=0$ -- we
   use the entire 
   cube in L0,  while in L1-L7, we use the central sphere of diameter
   $\sim0.8$ times the diameter of entire high-resolution region;
   column 3: mass of high-resolution particles; column 4: softening
   length of high-resolution particles; column 5: overdensity
   ($\delta_\mathrm{level} = \bar{\rho}_\mathrm{level} /
   \rho_{\mathrm{m}} - 1$) of the selected region at $z=0$. }
 \label{tab:table1}
 \begin{tabular}{ccccc}
  \hline
  level &  size/$\mathrm{Mpc}$ & $m_{\mathrm{p}}$/$\mathrm{M_{\odot}}$ & $\epsilon$/$\mathrm{kpc}$ & $\delta_{\mathrm{level}}$($z=0$) \\
  \hline

  L0 & $7.38 \times 10^{2}$  & $1.56 \times 10^{9}$  & $6.83$ & $0.0$ \\                  
  L1 & $8.12 \times 10^{1}$  & $7.41 \times 10^{5}$  & $5.31 \times 10^{-1}$ & $-0.607$ \\ 
  L2 & $1.23 \times 10^{1}$  & $1.45 \times 10^{3}$  & $5.61 \times 10^{-2}$ & $-0.918$ \\ 
  L3 & $1.65$                & $2.82$                & $8.32 \times 10^{-3}$ & $-0.964$ \\ 
  L4 & $2.22 \times 10^{-1}$ & $5.50 \times 10^{-3}$ & $1.04 \times 10^{-3}$ & $-0.974$ \\ 
  L5 & $4.55 \times 10^{-2}$ & $5.75 \times 10^{-5}$ & $2.27 \times 10^{-4}$ & $-0.976$ \\ 
  L6 & $9.43 \times 10^{-3}$ & $2.60 \times 10^{-7}$ & $3.77 \times 10^{-5}$ & $-0.986$ \\ 
  L7 & $1.58 \times 10^{-3}$ & $8.55 \times 10^{-10}$& $5.28 \times 10^{-6}$ & $-0.984$ \\ 

  \hline
 \end{tabular}
\end{table}

In the VVV, we select candidate resimulation regions to be nearly
spherical in shape and underdense (relative to the cosmic mean
density); initial conditions for these candidate regions are then
generated at higher resolution for each subsequent VVV resolution
level. This nested zoom technique is ideal to study extremely small
structures embedded within cosmologically representative environments
\citep{Jenkins2010,
  Jenkins2013,
  Jenkins2013b}. We refer the reader to \citet{Wang2020} for a
detailed description of the zoom-in strategy.
\cite{Wang2020} present 8 levels of resolution with uncut initial power
spectra\footnote{There are two additional levels (L7c and L8c) where
  the initial power spectrum is cut off on small scales to reflect 
  the free streaming of a 100 GeV neutralino. Simulations where this
  free streaming cutoff is resolved are known to produce spurious
  structures
  \citep[e.g.][]{Wang2007,Lovell2014}. Thus, for simplicity, we exclude these two
  levels from our analysis. }, labelled L0-L7. L0 refers to the
periodic, `parent' simulation cube with
$L_{\mathrm{box}}=738\ \mathrm{Mpc}$. Following \citet{Wang2020}, in
L1-L7 we only consider the halo population contained within 0.8 times
the radius of the high-resolution region, so as to avoid any potential
contamination from low-resolution particles in the boundaries of the
high-resolution regions. Details of the simulations at each resolution
level are given in Table~\ref{tab:table1}.

Haloes were identified using a friends-of-friends (FOF) algorithm
\citep[][assuming a linking length, $b=0.2$ times the mean
interparticle separation]{Davis1985}. Subhaloes within haloes were
identified using the SUBFIND algorithm \citep{Springel2001,
  Dolag2009}; both methods are built into \textsc{Gadget-4}. There are
many different definitions of halo mass: a basic approach is to adopt
the mass of the FOF group ($M_{\mathrm{FOF}}$) directly, while others
may use $M_{\Delta}$, defined as the mass scale within which the
average density is equal to $\Delta$ times the critical or the mean
density of the universe. The overdensity, $\Delta$, may be set in
various ways, e.g. $200$, $200\Omega_\mathrm{m}$ or $\Delta_\nu$ from
the spherical top-hat collapse model \citep{Eke1996,
  Bryan&Norman1998}. There is no consensus as to which definition is
`best'. \citet{Warren2006} argued that the FOF mass could suffer from
a systematic bias for haloes with small particle numbers, while
\citet{Tinker2008} calibrated the parameters in a fitting formula with
different mass definitions. In this paper, we adopt
$M_{\Delta=200\Omega_\mathrm{m}}$ ($M_{200}$ hereafter), as the
definition of halo mass. We consider only central haloes (excluding
subhaloes) in the following analysis.

\section{Results}
\label{sec:results}
\subsection{Halo mass function at $z=0$}
\label{sec:z0}

We begin by considering the halo mass function at $z=0$, shown in
Fig.~\ref{fig:fig1}, for each of the different resolution levels (L0-L7, solid
lines of different colours). We only include haloes containing at least
50 particles. We compare with the corresponding EPS prediction (dashed
colour  lines) based on the local overdensity and total mass of each
high resolution region. Error bars in the halo mass function are 
Poisson errors measured in each mass bin; these are usually largest at 
the high mass end because of the small number of high-mass haloes at each
resimulation level (e.g. only 18 for the largest mass bin of L0).

It can be seen that at $z=0$, EPS gives relatively more precise
predictions for the results of the simulations (within $\sim$ 20\%) at
all resolution levels compared to PS (black dashed line). This is
especially true at the higher levels (i.e., lower underdensity
regions). ST (gray dashed line)
gives the best prediction at L0 (the only at cosmic mean density),
resulting in the similarity between VVV/EPS (i.e. the ratio between
the simulation halo mass function and the EPS prediction) and ST/PS
(i.e. the ratio between the ST and PS halo mass functions, gray dashed line truncated at $50m_\mathrm{p,\ L0}$ in the bottom panel) lines for
L0. 

Our results from PS and ST for L0 at $z=0$ agree with previous studies
\citep[e.g.][]{Jenkins2001, Reed2003, Reed2007, Yahagi2004, Lukic2007,
  Pillepich2010}, suggesting that the ST model is a better
approximation than PS for volumes simulated at mean density and low
redshift. This is true even when we use $M_{200}$ rather than
$M_{\mathrm{FOF}}$ as the definition of halo mass. For example,
\citet{Reed2003} studied the halo mass function in the mass range
$\sim 10^{10.2}-10^{14.2}~\mathrm{M_{\odot}}$\footnote{We convert the mass unit of the works referred in this paper from $h^{-1}\mathrm{M_\odot}$ to $\mathrm{M_\odot}$ for easy comparison. } and found that PS
overpredicts the abundance by around $20\%$ when
$M<10^{13.7}~\mathrm{M_{\odot}}$ at $z=0$; ST, on the other hand, performs comparatively better in
the same mass range -- this is consistent with our findings that the
halo mass function of L0 at the low mass end
($M \lesssim 10^{14}~\mathrm{M_\odot}$) aligns with the ST prediction
and is $\sim 20\%$ lower than PS prediction.  On the other hand, it is
worth noting that while the simulated halo mass function tracks the ST
prediction well at the high mass end, the PS model underpredicts the
halo mass function of L0 at
$M \gtrsim 5 \times 10^{14}~\mathrm{M_\odot}$. This is also
consistent with the results of \citet{White1993, Gross1998,
  Governato1999, Jenkins2001, Lukic2007, Pillepich2010}.

In the higher resolution level simulations (L1-L7, corresponding to
increasingly underdense regions), the EPS model performs significantly
better than either PS or ST. 
A simple comparison can be made with the halo 
mass function at the high mass end of the L1 volume in Fig. \ref{fig:fig1} 
-- even with the most moderate underdensity ($\delta=-0.607$), both PS 
and ST significantly overpredict the halo abundance, while EPS 
provides a relatively accurate prediction.
This is to be expected because these
simulations focus on regions far below the average density of the
universe. Indeed, it is known that the halo mass function depends
strongly on local overdensity \citep{Gao2005, Rubino-Martin2008,
  Crain2009, Faltenbacher2010, Tramonte2017}. For example,
\citet{Gao2005} find that at $z=49$, the halo mass function in a
region with $\delta_\mathrm{nl}=4.3$ has a larger amplitude than that
in a region with $\delta_\mathrm{nl}=2.8$ by a factor of $\sim
4$. Only the EPS model takes local environment into consideration when
predicting the abundance of haloes.  
The mass function of L6 (blue
line) is less aligned with other levels. This may be due to cosmic
variance (i.e. the scatter in halo abundance across volumes of the same size and matter density; we refer readers to Section \ref{sec:realizations} for a detailed illustration). 
We examine this by comparing the halo mass function in L6 and the
corresponding volume in L5 (ensuring the same non-linear overdensity)
and find good agreement, which excludes the possibility of problems
related to numerical convergence in L6.  In general, EPS provides
reasonably accurate predictions (particularly at the low mass end)
which overestimate our simulated halo mass functions by
$\sim 20 \%$. This might due to various reasons, for example, the ambiguous definition of a ``virialized'' halo in the original PS theory.

\subsection{The redshift evolution of the halo mass function}
\label{sec:z2-30}

After examining the accuracy of theoretical predictions of the halo
mass function at $z=0$, we now consider how well these models perform
at higher redshift.  In Fig.~\ref{fig:fig2}, we show the halo mass
functions at redshifts $\sim$ 2, 7.8 and 30. For L0, at low
redshift, the simulation result is in good agreement with the ST
prediction. With increasing redshift, however, better agreement is
gradually found with the PS and EPS models (the PS and EPS predictions
are rather similar when $\delta=0$ and the enclosed mass is large
enough).

Our results at $z \lesssim 2$ are consistent with many previous
studies \citep{Reed2003, Yahagi2004, Lukic2007, Klypin2011}, which
find that the ST model is a good fit at low redshift. However, as
redshift increases, there are discrepancies among these simulation
studies: \citet{Reed2003} ($\sim10^{10.2-12.2}~\mathrm{M_\odot}$),
\citet{Lukic2007} ($\sim10^{8.2-10.2}~\mathrm{M_\odot}$), and
\citet{Wang2022} ($\sim10^{8.2-11.2}~\mathrm{M_\odot}$) suggested
that the deviation of ST from the simulation data (using $M_\mathrm{FOF}$ as the
halo mass) is $\lesssim 15\%$ at $z \lesssim 10$. On the other hand,
\citet{Hellwing2016} (also using $M_\mathrm{FOF}$) showed that ST
significantly overpredicts the halo abundance at $0.5 \le z \le 5$,
eventually becoming a better approximation at $z \sim
9$.

\citet{Cohn&White2008} discussed the differences arising from using
different definitions of halo mass at $z=10$: the halo mass function
with $M_\mathrm{FOF}$ in the mass bin
$10^{8.2-10.7}~\mathrm{M_\odot}$ agrees with the ST prediction at
$z=10$, while its counterpart with $M_{180}$ is almost half of the
$M_\mathrm{FOF}$ measurement. \citet{Klypin2011} (using
$M_{\Delta_\nu}$, with $\Delta_\nu$ the value from the spherical
top-hat collapse model) found that the deviation between ST and the
simulations becomes larger at higher redshift -- up to a factor of 10
in mass bin $10^{10.2-11.2}~\mathrm{M_\odot}$ at $z=10$. When using
a similar definition of halo mass, our results agree with the last two
studies, suggesting that the halo mass function with mass defined as
$M_{200}$ is lower than the ST prediction and approaches the PS or EPS
prediction at high redshift\footnote{As shown in Appendix
  \ref{ap:eagle34}, we corroborate our results with the EAGLE dark
  matter only simulations.}. We note that, when defined using
$M_\mathrm{FOF}$, the halo mass function (not shown) of L0 deviates
less from the ST prediction at high redshifts, which agrees with the
conclusion of \citet{Reed2003} that the halo mass function obtained
using spherical overdensity masses is lower than the mass function
based on the FOF mass.

\begin{figure}
    \centering
	\includegraphics[width=1.0\columnwidth]{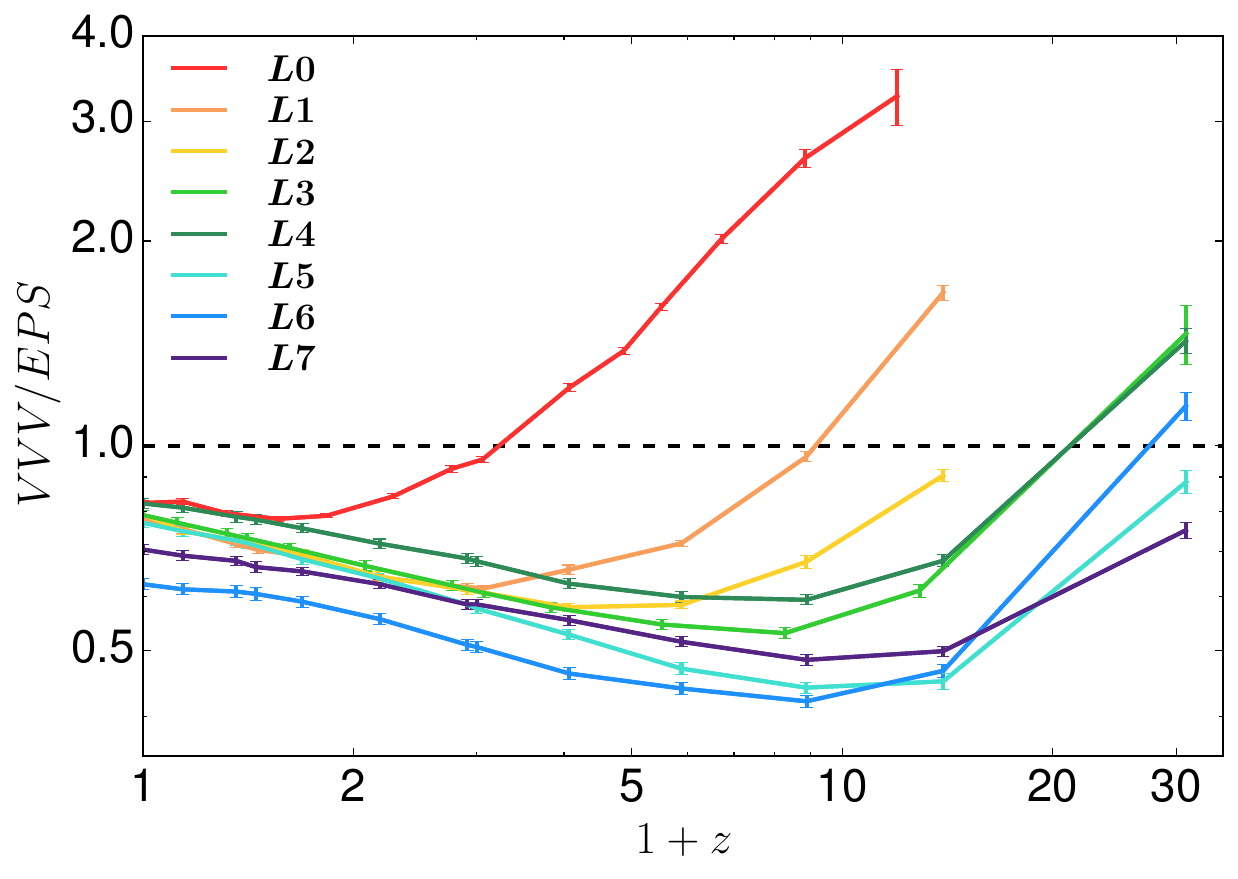}
    \vspace{-0.50cm}
	
	\caption{Evolution of the geometric mean of the ratio of the
          mass functions, VVV/EPS, in different resolution levels represented by
          the different colours. The error bars are the error of the 
          mean value obtained by error propagation; see the main text for
          details. 
          The deviations peak at $z \sim 2-9$ (depending on scale),
          indicating that the EPS predictions fit the simulations best
          at low and extremely high redshifts.  }
	\vspace{-0.20cm}
	\label{fig:fig3}
\end{figure}

\begin{figure*}
    \centering
    \includegraphics[width=2.0\columnwidth]{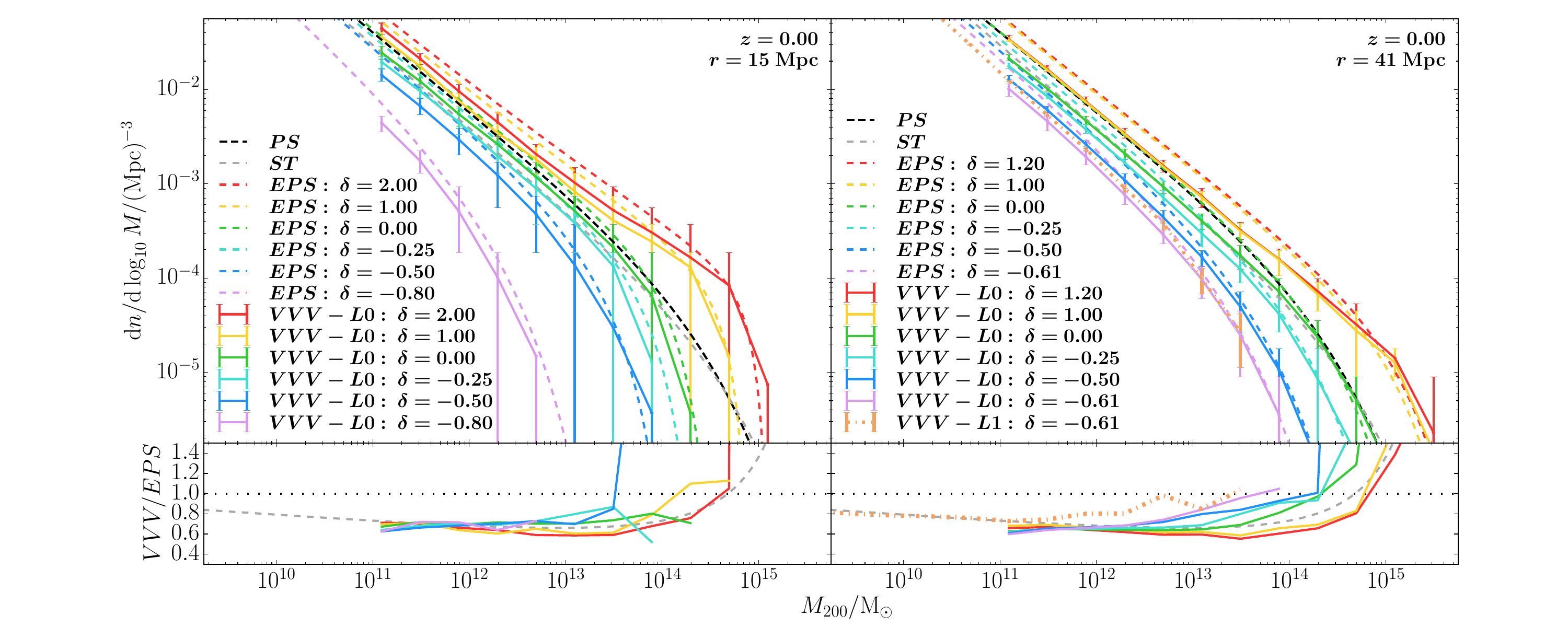}
    \includegraphics[width=2.0\columnwidth]{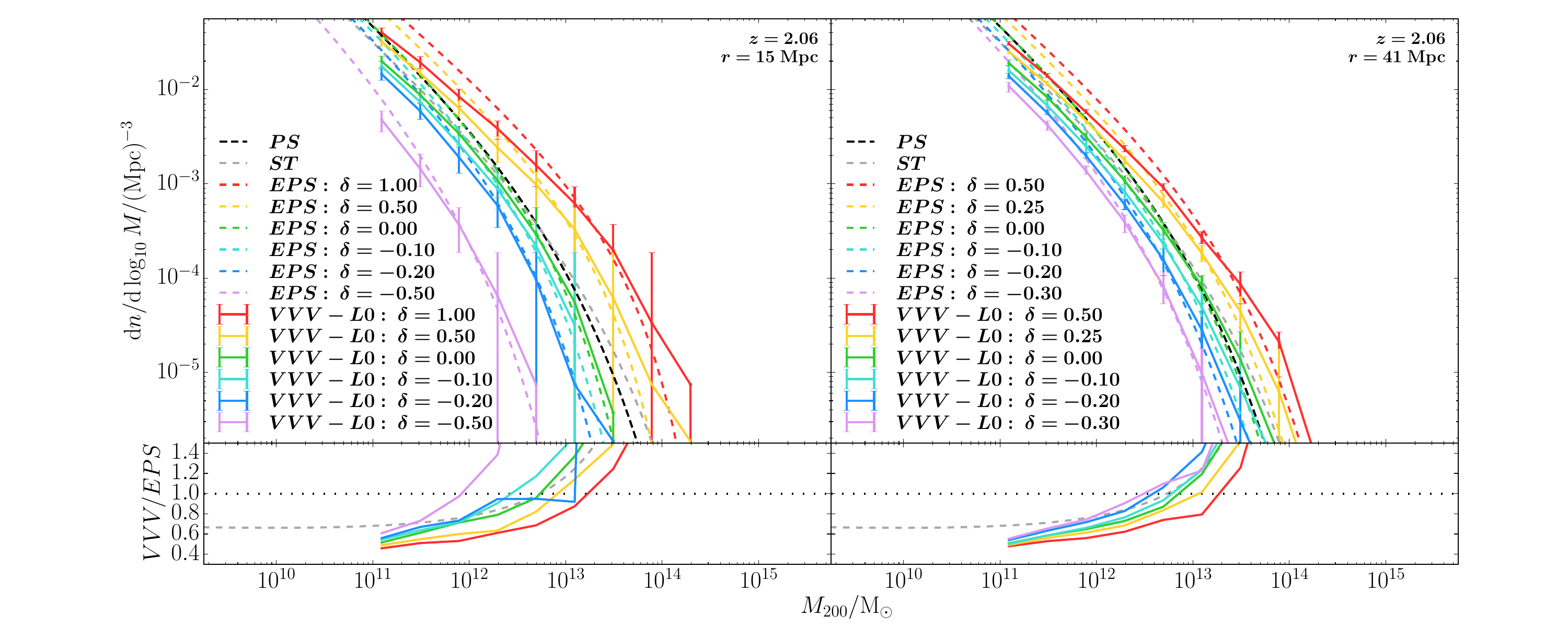}
    \includegraphics[width=2.0\columnwidth]{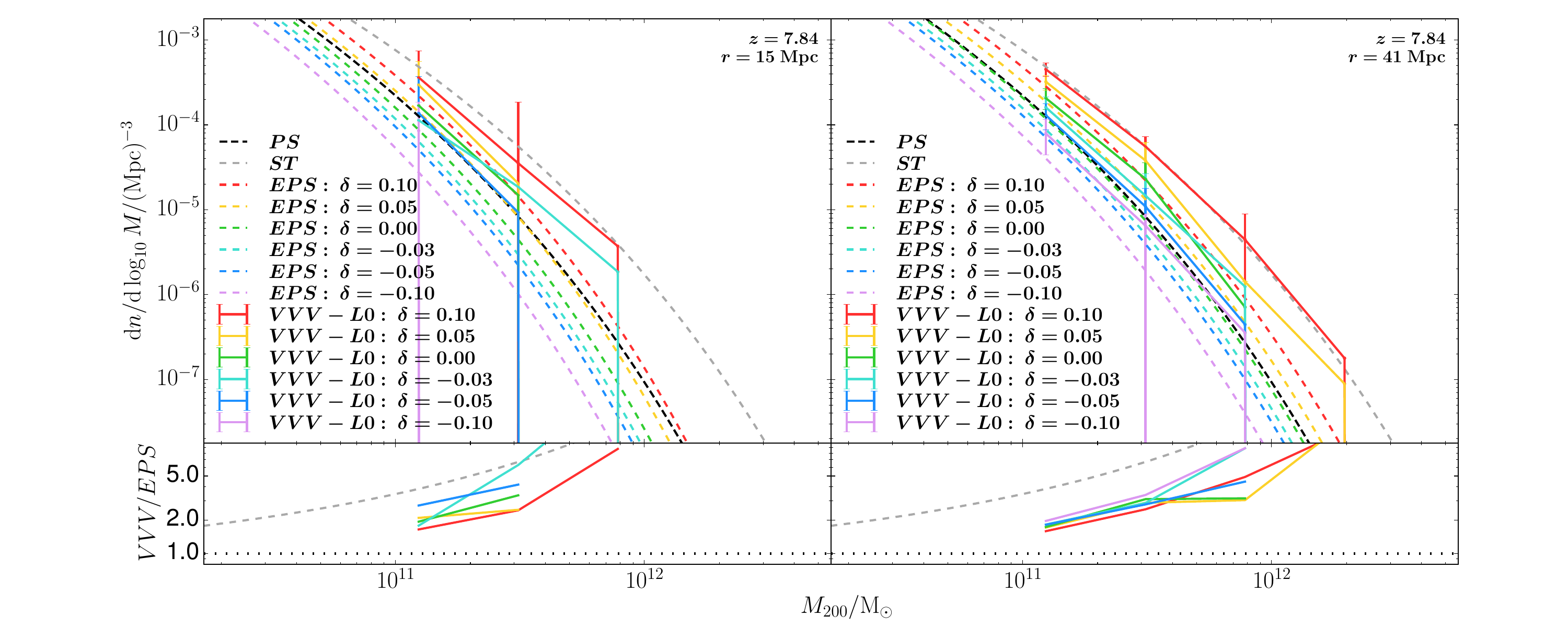}
    \vspace{-0.20cm}
    \caption{Average halo mass function in spherical regions of
      different overdensity selected from the L0 cube. The left and
      right columns correspond to different region sizes; each row
      corresponds to a different redshift (increasing from top to
      bottom). Black, gray and colour dashed lines show the PS, ST and
      EPS predictions respectively; the colour solid lines show the
      simulation results, with different colours corresponding to
      different overdensities. The error bars represent the 16-84th
      range amongst the regions; the orange dash-dotted line in the
      upper right panel shows the VVV L1 simulation with Poisson
      errors.  The halo mass functions of the different realizations
      are generally consistent with those obtained from the full
      simulation.  }
	
    \vspace{-0.30cm}
    \label{fig:fig4}
\end{figure*}

For the higher VVV resolution levels, the results at $z \sim 2$ are
quite similar to the results at $z=0$: the halo mass functions in the
simulations are still in fairly good agreement with the EPS
prediction. On the other hand, at $z \sim 7.84$ the difference between
the simulations and the EPS prediction becomes larger, with the ratio
of VVV/EPS dropping to $\sim 50\%$ for L2-L7. At $z=30$ results from
L3-L7 show that the low-mass end of the halo mass function (computed
so that haloes contain at least 50 particles and each bin contains at
least 20 halo samples) goes back to being aligned with EPS. 

In Fig.~\ref{fig:fig3}, we plot the evolution of the geometric mean of
the ratio VVV/EPS\footnote{We only use haloes with at least 50
  particles and mass bins with at least 20 samples. We checked that
  our results are independent of bin size and the precise threshold
  number of samples in each bin.} in each resolution level.
Error bars have been propagated from the Poisson errors of selected
mass bins using the following relations:
\begin{equation}
    g = \left(\prod^N_i{x_i}\right)^{1/N},\ \sigma_g = \frac{g}{N}\sqrt{\sum^N_i{\left(\frac{\sigma_i}{x_i}\right)^2}} ,
\end{equation}
where $g$ and $\sigma_g$ represent the value and the error of the
geometric mean ratio respectively; $x_i$ and $\sigma_i$ represent the
value of the ratio and the Poisson error in the $i$-th mass bin; and
$N$ is the number of mass bins.

Fig.~\ref{fig:fig3} suggests that the ratios in all resolution levels
follow similar tracks with time: the average VVV/EPS ratios are close
to one at the present, drop at intermediate redshifts, and increase
again at earlier times. Generally, the higher the level is, the
earlier their track intersects the dashed horizontal line at
unity, which is likely related to the different characteristic mass
scales of haloes in each level. 

We find that the average VVV/EPS ratios are close to unity at low
redshifts for large masses and, at high redshifts, for low
masses. This is consistent with the results of \citet{Gao2005}, who
studied the halo mass function in denser environments at $z=0$ and
$z=49$, and showed that EPS gives accurate predictions in the mass
ranges $10^{8.2-10.7}~\mathrm{M_\odot}$ ($z=0$), and
$10^{2.7-5.2}~\mathrm{M_\odot}$ ($z=49$), confirming that EPS is
a good fit at low redshift and extremely high redshift.

\subsection{Halo mass functions with different realizations}
\label{sec:realizations}

The analyses in the preceding section use only one realization for
each resolution level and are thus subject to ``cosmic variance''. To
assess the effect of this variance and test the accuracy of the
theoretical models of the halo mass function in general, we picked
spherical regions of different sizes and overdensities from the L0
cube, and measured the average halo mass function in them. The L0
volume is large enough to sample a range of overdensities with
statistical fidelity. We generated $10^{5}$ randomly located points and
measured their local overdensities inside spheres of different radii,
and then selected 100 samples with the closest overdensity to a set of
preselected values. We then computed the mean halo mass function
across these 100 samples; this is displayed in Fig.~\ref{fig:fig4}.
Black (gray or colored) dashed lines represent the predictions of PS
(ST or EPS), and the colour solid lines show the mean halo mass
function of samples in the VVV simulation.  Different colors represent
different overdensities, the values of which are adapted to the
redshift and sphere size of each panel (marked at the upper right
corner), in order to obtain a large enough sample.

\begin{figure}
    \centering
    \includegraphics[width=1.0\columnwidth]{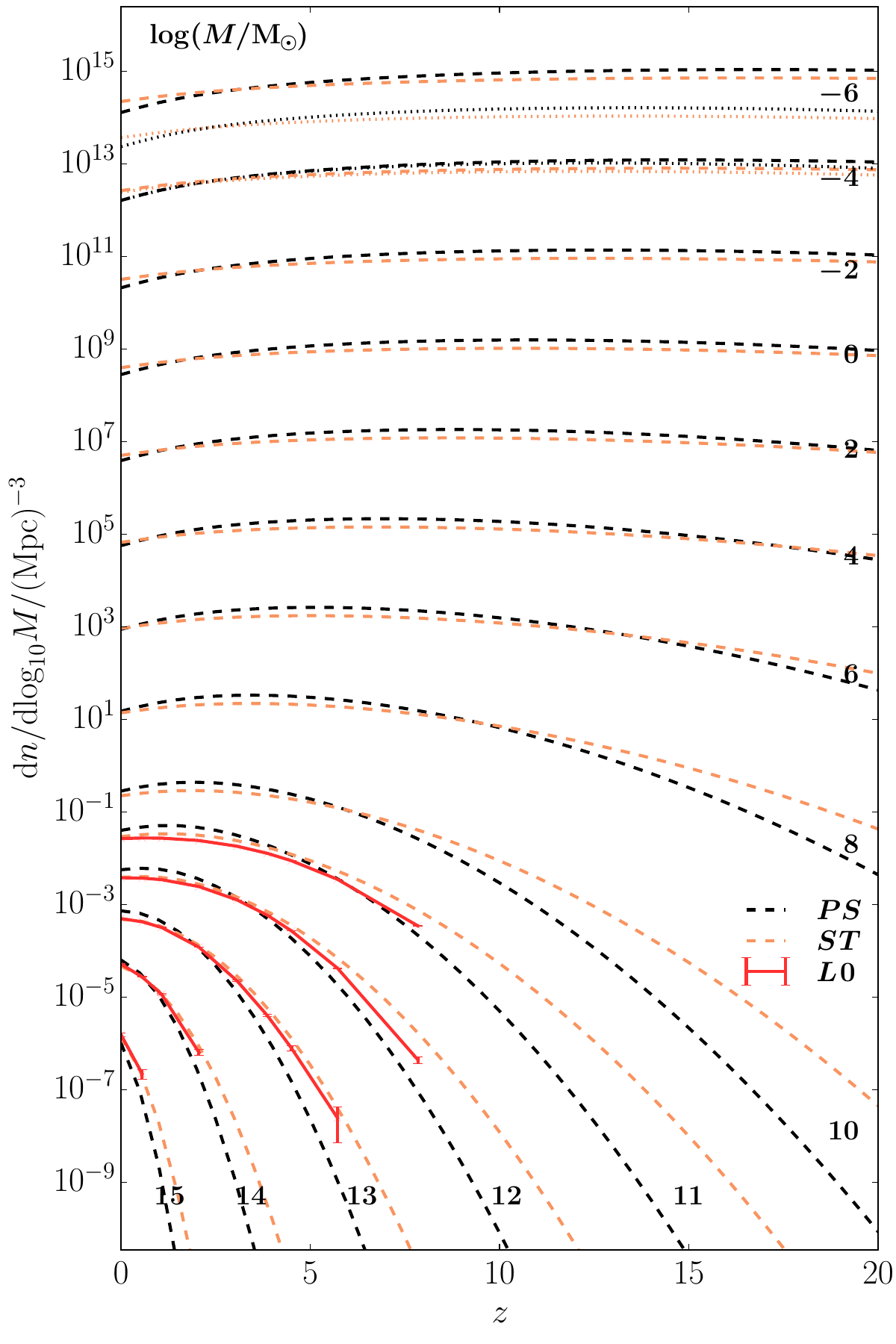}
    \vspace{-0.40cm}
    \caption{The evolution of halo abundance within
        the mass range $10^{-6}$--$10^{15}~\mathrm{M_\odot}$ in the
        whole universe (or mean density regions) predicted by PS (black
        lines) and ST (coral lines), with the numbers beside each line
        representing the corresponding $\log(M/\mathrm{M_\odot})$. The
        dashed lines show the prediction with the uncut initial power
        spectrum, while the dotted lines take account of the
        free-streaming effect at
        $10^{-6}$--$10^{-4}~\mathrm{M_\odot}$. The red lines are the
        halo mass function in the L0 simulation, with the error bars
        representing Poisson errors. Within the mass range
        $10^{11}$--$10^{15}~\mathrm{M_\odot}$, ST is a good
        approximation at low redshift $z \lesssim 2$, but overpredicts
        the halo number density at earlier times. The free-streaming
        effect suppresses the abundance of $10^{-6}~\mathrm{M_\odot}$
        haloes by approximately one order of magnitude, while barely
        affecting larger haloes. }
    \vspace{-0.38cm}
    \label{fig:fig5}
\end{figure}

It can be seen that the ratio $\mathrm{VVV/EPS}$ (for $M_{200}$)
depends only weakly on sphere size and overdensity, but evolves with
redshift -- it agrees with the theoretical ST/PS line at $z=0$ and
$2$, but deviates at $z=7.84$.  The mean halo mass function averaged
over the random samples is quite similar to that of the full volume,
corroborating our conclusion from the full simulation that the ST
model provides a good approximation to the halo mass function at low
redshifts, but overpredicts it at high redshifts.  However, the size
of the error bars suggests that even for regions with the same
overdensity and total mass, there is considerable variance in the halo
mass function. This is especially true in smaller regions and might
be part of the reason why the predicted halo abundance in some levels (e.g. L6) are less accurate compared to other levels.
In the upper right panel of Fig.~\ref{fig:fig4}, we also show the halo
mass function of L1 at $z=0$ ($r=41\ \mathrm{Mpc}$, $\delta=-0.61$)
with an orange dash-dotted line. 
Comparing to its counterpart in L0 (purple solid line), we find that
the L1 mass function is a little larger, by $\lesssim 10\%$, at
$M_{200} \sim 10^{12.5}~\mathrm{M_\odot}$, showing reasonable
convergence in the halo mass function between the two different levels.

\section{Discussion}
\label{sec:discussion}
In the previous section, we have shown that for different subvolumes of the L0
  simulation (corresponding to regions with a range of overdensities), the halo abundance in the EPS model differs from the   
  halo abundance in the simulation at a similar level (20 - 40\% at $z=0$, higher at high $z$) as 
  in the L0 full volume, which has cosmic mean density.
  We expect this 
  conclusion to hold not just for a relatively small range in halo mass, 
 but also for smaller masses, down to the cutoff in the power spectrum; we are
  unable to perform similar tests for higher levels due to the limited
  sample size in zoomed regions.  Consequently, the deviations between the EPS predictions and the (L1 - L7) VVV simulations (Fig.~\ref{fig:fig1} and Fig.~\ref{fig:fig2}) should
  approximately reflect the difference at other overdensities, rather than just the
  underdense regions probed by the VVV simulations. 
Combining these two arguments, given that the PS formalism is equivalent to integrating the EPS formula over all linear overdensities, $\delta_0$, we can use the PS formula to predict the abundance of haloes across the entire range of halo masses at the mean density. 

Based on this, in Fig.~\ref{fig:fig5}, we present the
  evolution of halo abundance in the whole universe (at mean density)
  as predicted by the PS (black) and ST (coral) formalisms. The predictions are similar to
  those of \citet{Mo2002}, but we show the {\it differential} mass
  function, and display it over a much more extended mass range than 
  \citet{Mo2002} and with updated cosmological parameters.
The comparison with the halo mass function in the L0 simulation (red
lines) corroborates our previous results that, at low redshift ($z
\lesssim 2$) and high masses ($10^{11-15}~\mathrm{M_\odot}$), the ST
model provides a better prediction, while the PS model underpredicts the
abundance of large haloes ($\gtrsim 10^{15}~\mathrm{M_\odot}$) and
overpredicts the abundance of smaller haloes ($\sim
10^{11-14}~\mathrm{M_\odot}$). At higher redshifts, the simulation
results lie between the PS and ST predictions. 

For the two lowest mass bins
  (i.e. $10^{-6}~\mathrm{M_\odot}$ and $10^{-4}~\mathrm{M_\odot}$), we
  show the effect of the free-streaming cut-off for the 100 GeV
  neutralino with dotted lines, using the cut-off initial power
  spectrum of \citet{Wang2020}. Note that here we have used the sharp
  $k$-space window function to evaluate $\sigma(M)$ in the theoretical
  mass functions, as suggested by \citet{Benson2013}. This
  modification provides a more accurate prediction of the halo
  abundance near the cut-off scale compared to a real-space top-hat
  filter, which instead overpredicts the mass function near the
  cut-off by re-weighting long wavelength modes.  We find that the
  effect of free-streaming is to suppress the abundance of
  $10^{-6}~\mathrm{M_\odot}$ haloes by one order-of-magnitude, while
  larger haloes are barely affected\footnote{We also
      compared the predictions with the same sharp $k$-space filter,
      but different initial power spectra, and confirmed that the
      suppression comes from the free-streaming effect. }.  This is
  consistent with previous studies of warm dark matter models
  \citep[e.g.][]{Schneider2012, Benson2013, Bose2016}, which suggest
  that free-streaming only suppresses the halo mass function near and
  below the cut-off scale.  

\section{Summary and conclusions}
\label{sec:conclusion}
In this paper, we have made use of the VVV simulations, a suite of
multi-zoom nested simulations at very high resolution, to test the
accuracy of the Press-Schechter (PS), Sheth-Tormen (ST) and extended
Press-Schechter (EPS) models for the abundance of haloes as a function
of their mass and redshift. In particular, this work focuses on the
halo mass function at extremely small scales (down to
$\sim 10^{-6}~\mathrm{M_\odot}$) and its evolution to high redshift
($z=30$). The resolution levels are labelled L0-L7, corresponding to
smaller, higher density regions of progressively higher mean underdensity
(below the cosmic mean). 
The non-linear underdensity of each level is mapped into a linear
underdensity using fitting formulas from the EPS formalism.

We find that at $z=0$, the ST model provides the most accurate fit to
the halo mass function in the L0 volume (overdensity, $\delta=0$;
$10^{11-15}~\mathrm{M_\odot}$) but the EPS model provides the best
fit to the higher resolution levels (L1-L7; $\delta<-0.6$ at
$10^{-6-12.5}~\mathrm{M_\odot}$), to better than $20\%$ accuracy (see
Fig.~\ref{fig:fig1}). The results at $z=2$ are similar, but there are
larger deviations at $z \sim 7-15$ from the ST prediction in L0 and
from the EPS predictions in the higher resolution levels. However, at
even higher redshift, $z \sim 30$
($10^{-6-2.5}~\mathrm{M_\odot}$; $-0.55<\delta<-0.30$), the EPS model provides, once more, a good fit to the halo
mass functions in the simulations (see Fig.~\ref{fig:fig2} and
Fig.~\ref{fig:fig3}).  We validated our results by selecting
regions of different volume and overdensity from the full L0
volume, and find that the VVV/EPS ratio (for $M_{200}$)
depends only weakly on region size and overdensity, but
increasingly deviates from unity at higher redshifts
(Fig.~\ref{fig:fig4}). Finally, we tested convergence by comparing the
halo mass function in L1 with those in selected subvolumes of L0
spanning a range of sizes and overdensity.

Having demonstrated that the EPS formalism gives a good
description of the halo mass function in the biased regions of the VVV
high resolution levels, given the arguments in Section~\ref{sec:discussion}, it is reasonable to assume that the PS formalism (or the ST formula)
also gives a good description of the halo mass function in
representative, mean density regions. 
Thus, we are able to present the
actual halo mass function (number of haloes per unit volume) over the
entire mass range in $\Lambda$CDM, from $10^{-6}$ to
$10^{15}~\mathrm{M_\odot}$ (Fig.~\ref{fig:fig5}). While the mass function at the high mass end
is, of course, well known, our study is the first to explore the mini-halo
regime ($\lesssim 10^5~\mathrm{M_\odot}$) where the EPS model provides
a prediction for the halo mass function, with deviations at the
$\sim 20-50\%$ level, which could be reduced with further theoretical
work.  Finally, we note that since we have analyzed dark matter only
simulations, the impact of baryons on the haloes is ignored.  A more
complete study based on hydrodynamics simulations with full baryon
physics will be presented in a forthcoming paper.

\section*{Acknowledgements}
We thank the anonymous referee for a constructive 
report that helped us tighten up our analytical arguments and improve our manuscript. 
We would like to thank Simon D. M. White, Volker Springel,
and Shaun Cole for useful discussions and advice. 
We acknowledge support from the National Natural Science
Foundation of China (Grant No. 11988101, 11903043, 12073002,
11721303) and the K. C. Wong Education Foundation. 
HZ acknowledges support from the China Scholarships 
Council (No. 202104910325). SB is supported by the UK
Research and Innovation (UKRI) Future Leaders Fellowship [grant number
MR/V023381/1]. CSF acknowledges support by the European Research
Council (ERC) through Advanced Investigator grant, DMIDAS (GA
786910). 
JW acknowledges the support of the
research grants from the Ministry of Science and Technology
of the People’s Republic of China (No. 2022YFA1602901), 
the China Manned Space Project (No. CMS-CSST-2021-B02), 
and the CAS Project for Young Scientists in Basic Research (Grant No. YSBR-062). 
This work used the DiRAC@Durham facility managed by the
Institute for Computational Cosmology on behalf of the STFC DiRAC HPC
Facility (www.dirac.ac.uk). The equipment was funded by BEIS capital
funding via STFC capital grants ST/K00042X/1, ST/P002293/1,
ST/R002371/1 and ST/S002502/1, Durham University and STFC operations
grant ST/R000832/1. DiRAC is part of the UK National e-Infrastructure.

\section*{Data availability}
The data used in this paper will be shared upon reasonable request
to the corresponding author. 

\bibliographystyle{mnras}
\bibliography{main} %

\appendix

\section{Conversion between linear and non-linear overdensity}
\label{ap:eq6}

\begin{figure}
    \centering
	\includegraphics[width=1.0\columnwidth]{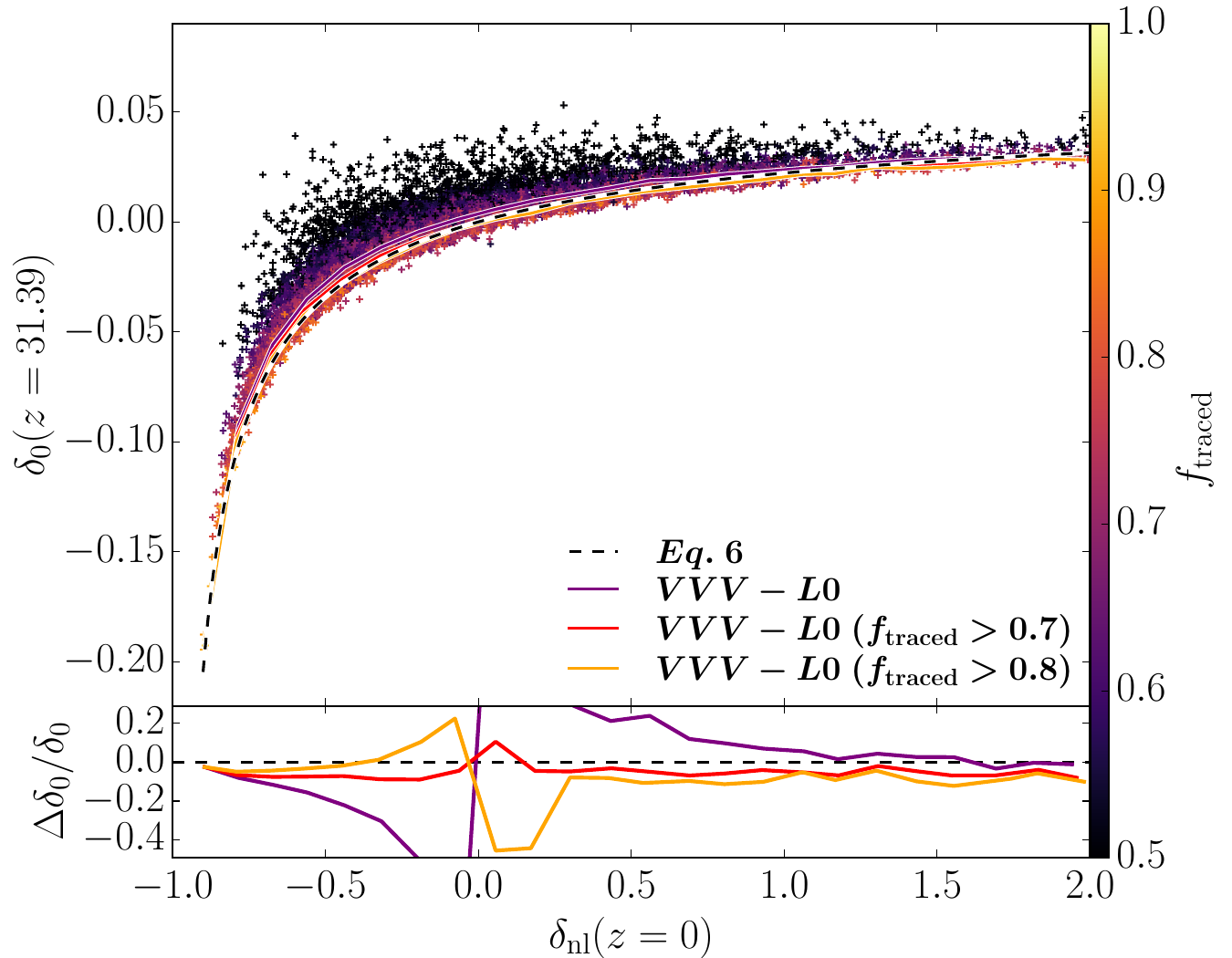}
	\vspace{-0.40cm}
	\caption{Conversion between non-linear and linear
          overdensities. The colour of each cross represents the
          fraction of traced particles in the volume at $z=31.39$,
          indicating the extent to which the measurement of
          overdensity in that volume is contaminated by background
          particles; see the text for details. The purple solid line
          shows the median values for all samples and the red and
          orange lines the median values of samples with
          $f_\mathrm{traced}>0.7$ and $0.8$ respectively. The black
          dashed line shows the prediction of Eq.~\ref{eq:6}. The
          bottom panel shows the ratio of the difference relative to
          the predicted value. Eq.~\ref{eq:6} provides a reasonably
          accurate fit to the simulations, especially for samples with
          high values of $f_\mathrm{traced}$.}
	
	\label{fig:figA1}
\end{figure}

\begin{figure*}
    \centering
	\includegraphics[width=2.0\columnwidth]{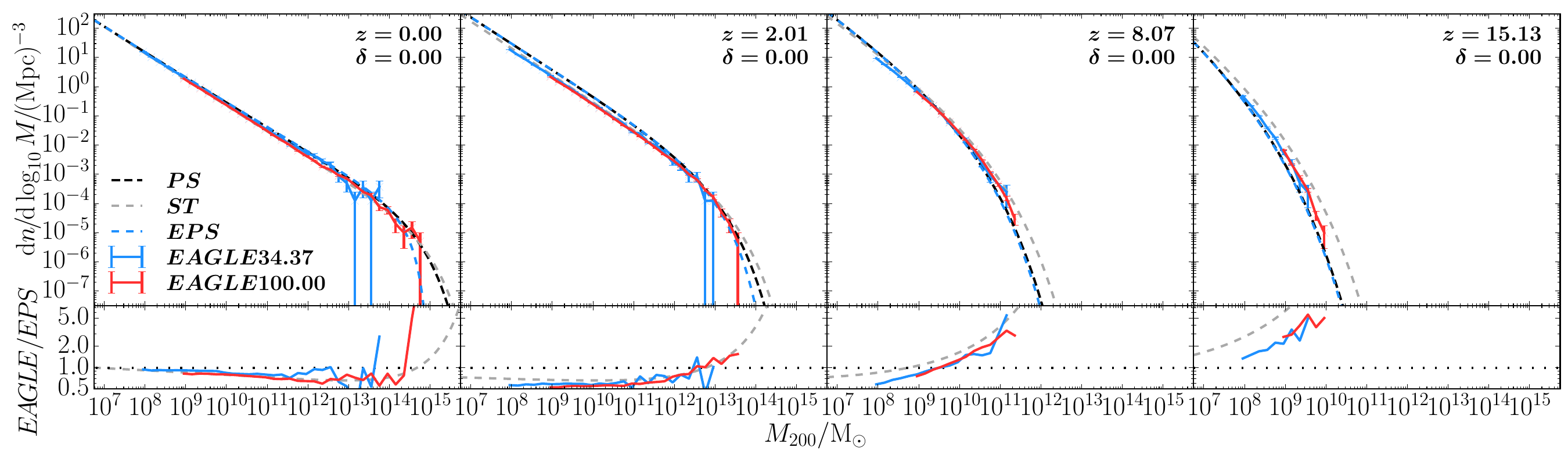}
	\vspace{-0.30cm}
	\caption{Comparison of the PS, ST and EPS models with the
          halo mass functions measured in dark matter-only versions of
          the EAGLE simulation at different redshifts. 
          The colour solid lines show the halo mass function in the EAGLE
          simulations of  side $L_\mathrm{box}=34.37\
          \mathrm{Mpc}$ (blue) and $100.00\ \mathrm{Mpc}$ (red); 
          the error bars are Poisson errors. Black, gray, and
          colour lines show the predictions of the PS, ST, and
          EPS models. The EPS predictions were
          calculated for the entire mass in the EAGLE34.37 cube; for the larger total mass
    in the EAGLE100.00 box, the EPS predictions almost overlap with the PS
    model and are not shown for clarity. 
    In the bottom panels, the colour solid lines show the ratio,
    EAGLE/EPS, while black dotted lines indicate a value if one; the gray
    dashed lines represent ST/PS.  The results from the EAGLE
    simulations align well with our results from the VVV
    simulations. } 
	
	\label{fig:figA2}
\end{figure*}

We converted the non-linear overdensity measured in the simulation to
a linear overdensity using Eq.~\ref{eq:6}, as required by the EPS
theory. Here, we examine whether this fitting formula is still
accurate for the extremely underdense regions of interest in VVV.  We
generated $10^4$ random positions in the VVV-L0 volume at $z=0$, and
measured the non-linear overdensities, $\delta_\mathrm{nl}(z=0)$,
within spheres of radius $r=15\ \mathrm{Mpc}$, each of which would
contain $\sim 10^6$ particles if $\delta_\mathrm{nl}=0$.  For each
sphere, we then traced the contained particles back to $z=31.39$. We
then constructed the polyhedron with the smallest volume containing
all these particles. The polyhedron was used to estimate the
non-linear overdensity, $\delta_\mathrm{nl}(z=31.39)$\footnote{ This
  method neglects the edges between the particle spheres and the
  boundary of the region, but since the particle number is large
  enough, it only overestimates the density by $\lesssim 3\%$. }. We
tested Eq.~\ref{eq:6} by converting the two measured non-linear
overdensities into linear overdensities\footnote{It is worth noting
  that at such high redshift ($z=31.39$), $\delta_0$ is very close to
  $\delta_\mathrm{nl}$ for most regions as structures have not yet
  formed and $\delta_\mathrm{nl}(z=31.39) \sim 0$. However, the
  difference increases as $\delta_0 \rightarrow -\infty$ with
  $\delta_\mathrm{nl} \rightarrow - 1$ according to the formula. } and 
comparing them.

We note that it is very likely for the polyhedron at $z=31.39$ to
contain particles that no longer belong to the corresponding sphere at
$z=0$. To account for this, we define $f_\mathrm{traced}$ as the
fraction of particles that are traced from $z=0$ to the polyhedron,
describing the extent to which the measurement of the overdensity in this volume
is uncontaminated by other particles. In other words, a low value of
$f_\mathrm{traced}$ means that the polyhedron at $z=31.39$ contains a
small fraction of particles that eventually end up in the
corresponding sphere at $z=0$.

In Fig.~\ref{fig:figA1}, we plot the linear overdensity at $z=31.39$
against the non-linear overdensity at $z=0$. The bottom panel shows
the ratio between the difference and the predicted value. The purple
line, representing the median value for all samples (including those
highly contaminated volumes with low $f_\mathrm{traced}$), deviates
most from the black dashed line, which is the prediction from
Eq.~\ref{eq:6}. The red and orange lines are restricted in
$f_\mathrm{traced}$ (indicating uncontaminated volumes); these
predictions are well aligned with the results in simulation. The
largest values of $|\Delta \delta_0/\delta_0|$ occur near the middle
of the range, because $\delta_0$ itself is around zero as
$\delta_\mathrm{nl} \sim 0$.
We also tested Eq.~\ref{eq:6} with the correction function, $C=1$, 
but the difference is negligible since the lowest non-linear overdensity here is only $\sim-0.9$
due to a lack of samples, where the \citet{Mo1996} analytical fit and numerical solution are still very similar.

\section{Tests in full-box simulations with higher resolution}
\label{ap:eagle34}

We also compared the predictions of the different models with the halo
mass functions in dark-matter-only versions of the EAGLE simulation
\citep{Schaye2015} with the same power spectrum but higher resolution
than VVV-L0 (EAGLE34.37:
$L_\mathrm{box}=34.37\ \mathrm{Mpc}, N=1034^3,
m_\mathrm{dm}=1.44\times 10^5~\mathrm{M_\odot}$, and EAGLE100.00:
$L_\mathrm{box}=100.00\ \mathrm{Mpc}, N=1504^3,
m_\mathrm{p}=1.15\times 10^7~\mathrm{M_\odot}$). The results are
quite similar to those for VVV-L0: the ST model provides a good
description of the halo mass functions in the simulations at low
redshifts, but overestimates them at high redshifts ($z \gtrsim 8$),
especially at the high mass end. Overall, the models provide reasonably 
accurate predictions that align with the results in the main text.

\bsp	
\label{lastpage}
\end{document}